\pdfoutput=1

\documentclass[prb,twocolumn,showpacs,showkeys,preprintnumbers,superscriptaddress,floatfix]{revtex4-2}
\usepackage[T1]{fontenc}
\usepackage[utf8]{inputenc}
\inputencoding{latin1}
\usepackage{amsbsy}
\usepackage{amstext}
\usepackage{graphicx}

\usepackage[english]{babel}
\usepackage{amsmath}
\usepackage{amsfonts}
\usepackage{color}
\usepackage{amssymb}
\usepackage{array}
\usepackage{makecell}
\usepackage{makeidx}
\usepackage{float}
\usepackage{multirow}
\usepackage{times}

\usepackage{wasysym}

\newcommand{\NaMnO}{(NaMn$_3$)Mn$_4$O$_{12}$}
\newcommand{\CaMnO}{(CaMn$_3$)Mn$_4$O$_{12}$}

\newcommand{\QP}{($AA'_3)B_4$O$_{12}$}

\begin{document}

\title{Direct observation of Jahn-Teller critical dynamics at a charge-order Verwey transition}

\author{Vin\'icius Pascotto Gastaldo}
\email{vinicius.gastaldo@ufms.br}
\affiliation{Departamento de F\'isica, Universidade Federal de S\~ao Carlos, S\~ao Carlos
- SP, 13565-905, Brazil}
\affiliation{IMPMC, Sorbonne Universit\'e and CNRS, 4, place Jussieu, 75005 Paris, France}
\affiliation{Instituto de F\'isica, Universidade Federal do Mato Grosso do Sul, Campo Grande - MS, 79070-900, Brazil}

\author{Mala N. Rao}
\affiliation{Solid State Physics Division, Bhabha Atomic Research Centre, Trombay, Mumbai 400085, India}
\affiliation{Homi Bhabha National Institute, Anushaktinagar, Mumbai 400094, India}

\author{Alexey Bosak}
\affiliation{European Synchrotron Radiation Facility, 6 rue Jules Horowitz, 38043
Grenoble, France}

\author{Matteo d'Astuto}
\affiliation{IMPMC, Sorbonne Universit\'e and CNRS, 4, place Jussieu, 75005 Paris, France}
\affiliation{Institut N\'eel CNRS/UGA UPR2940, 25, rue  des  Martyrs, 38042 Grenoble, France}

\author{Andrea Prodi}
\affiliation{European Synchrotron Radiation Facility, 6 rue Jules Horowitz, 38043
Grenoble, France}

\author{Marine Verseils}
\thanks{Present Address: Ligne AILES - Synchrotron SOLEIL, 91190 Gif-sur-Yvette CEDEX, France.} 
\affiliation{IMPMC, Sorbonne Universit\'e and CNRS, 4, place Jussieu, 75005 Paris, France}

\author{Yannick Klein}
\author{Christophe Bellin}
\affiliation{IMPMC, Sorbonne Universit\'e and CNRS, 4, place Jussieu, 75005 Paris, France}

\author{Luigi Paolasini}
\affiliation{European Synchrotron Radiation Facility, 6 rue Jules Horowitz, 38043
Grenoble, France}

\author{Adilson J. A. de Oliveira}
\affiliation{Departamento de F\'isica, Universidade Federal de S\~ao Carlos, S\~ao Carlos
- SP, 13565-905, Brazil}

\author{Edmondo Gilioli}
\affiliation{Istituto dei Materiali per Elettronica e Magnetismo, CNR, Area delle
Scienze, 43100 Parma, Italy}

\author{Samrath Lal Chaplot}
\affiliation{Solid State Physics Division, Bhabha Atomic Research Centre, Trombay, Mumbai 400085, India}
\affiliation{Homi Bhabha National Institute, Anushaktinagar, Mumbai 400094, India}

\author{Andrea Gauzzi}
\email{andrea.gauzzi@sorbonne-universite.fr}
\affiliation{IMPMC, Sorbonne Universit\'e and CNRS, 4, place Jussieu, 75005 Paris, France}

\date{\today}

\begin{abstract}
By means of diffuse and inelastic x-ray scattering (DS,IXS), we probe directly the charge-ordering (CO) dynamics in the Verwey system \NaMnO, where a peculiar quadruple perovskite structure with no oxygen disorder stabilizes a nearly full Mn$^{3+}$/Mn$^{4+}$ static charge order at $T_{\rm CO}$=175 K concomitant to a commensurate structural modulation with propagation vector ${\bf q}_{\rm CO}=(\frac{1}{2},\frac{1}{2},0)$. At $T_{\rm CO}$, the IXS spectra unveil a softening of a 35.3 meV phonon at ${\bf q}_{\rm CO}$. Lattice dynamical calculations enable us to attribute this soft phonon to a A$_g$ mode whose polarization matches the Jahn-Teller-like distortion pattern of the structural modulation. This result demonstrates that the Jahn-Teller instability is the driving force of the CO Verwey transition in \NaMnO, thus elucidating a long-standing controversy regarding the mechanism of this transition observed in other mixed-valence systems like magnetite.  
\end{abstract}

\maketitle

\section{Introduction}

The interplay of electron and lattice dynamics beyond the adiabatic approximation is at the forefront of research in the study of many-body systems, such as chemical reactions \cite{yar12}, ionization processes \cite{gal17} and polaron transport \cite{hua17}. Favorable conditions to investigate the non-adiabatic regime are usually found in mixed-valence systems, since charge fluctuations drive a strong electron-lattice coupling. One fascinating situation occurs when these fluctuations stabilize a static order of valence electrons. In a seminal paper \cite{ver39}, Verwey proposed that this situation is indeed realized as a static Fe$^{2+}$/Fe$^{3+}$ charge order (CO) in the archetypal mixed-valence system magnetite Fe$_{3}$O$_{4}$, which would explain a pronounced structural distortion concomitant to a jump of the electrical resistivity at $T_{V}=120$ K, known as Verwey transition. Verwey's picture has been subsequently questioned because a simple ionic description does not account for the complex structural changes occurring at $T_{V}$ \cite{wal02,gar04,shc09,wri01,wri02,roz06,roo18} and an alternative picture of polarons involving the dynamics of 3 Fe sites (trimerons) has been proposed \cite{sen12,hoe13}. Subsequent studies of the critical dynamics at the transition have given evidence of electronic \cite{bal20} but no phonon \cite{hoe13,bor20} mode softening, which leaves open the question of the cooperative role of the electronic and lattice degrees of freedom on the transition.      

A similar phenomenology is found in other mixed-valence systems like chemically doped manganites and nickelates with perovskite-like structure \cite{hot06,coe04,coe99}. Also in these systems, a static CO picture seems to be oversimplified due to disorder and electronic inhomogeneities inherent to chemically substituted compounds and alternative scenarios of dynamic charge fluctuations like Zener-polarons \cite{dao02} have been proposed. In order to elucidate the above controversial points, here we investigate the Jahn-Teller (JT) dynamics governing the interplay of charge, orbital and spin orderings in mixed-valence transition-metal compounds, as discussed in the seminal papers by Goodenough \cite{goo55} and by Wollan and Koehler \cite{wol55} and extensively studied later \cite{cha74,pin06,rad97}.
The challenge is that a reliable investigation of this dynamics is hindered in most mixed-valence systems by the coexistence of charge-ordered and -disordered phases \cite{rad97}, incommensurate structural modulations like stripe phases \cite{tra96} and electronic phase separation \cite{dag01}.


In the present work, we show that the above difficulty is overcome in the mixed-valence compound \NaMnO\ (NaMnO) \cite{mar73}. Similar to magnetite and manganites, at low temperatures, $T_{\rm CO}$=175 K, NaMnO exhibits a Mn$^{3+}$/Mn$^{4+}$ CO transition of the Verwey type consisting of a cubic $Im\bar{3}$ to monoclinic $I2/m$ structural transition accompanied by a large jump of the electrical resistivity \cite{pro04}. The unique characteristics of the quadruple perovskite structure \QP\ allow to obtain an equal proportion of Mn$^{3+}$ and Mn$^{4+}$ ions in the octahedral $B$-sites without chemical substitutions. In addition, the peculiar four-fold coordination of the Mn$^{3+}$ ions in the $A'$ sites avoids the formation of oxygen defects. It turns out that \NaMnO\ exhibits an almost full charge order of the $B$-site Mn$^{3+}$ and Mn$^{4+}$ ions concomitant to a zig-zag ordering of the $e_g$ 3$d_{z^2}$ orbitals of the Mn$^{3+}$ ions and leading to a comparatively simple structural modulation with commensurate propagation vector ${\bf q}_{\rm CO}=\left(\frac{1}{2},\frac{1}{2},0\right)$ \cite{pro14}.

The above favorable conditions prompt us to probe the lattice dynamics governing the CO Verwey transition in NaMnO. We first searched for structural anomalies at $T_{\rm CO}$ by diffuse x-ray scattering (DS), a suitable probe of static short-range structural correlations precursor of a structural transition. As reported on magnetite \cite{bos14}, these correlations produce anomalous DS intensities different from ordinary thermal diffuse scattering. Second, we investigated dynamic anomalies by inelastic x-ray scattering (IXS) at $T_{\rm CO}$. Owing to the low-$Z$ of Na, the scattered intensities mainly arise from the Mn and O atoms playing the dominant role in the lattice dynamics, a further favorable condition for both experiments.

\section{Experimental Details and Lattice dynamical calculations}

NaMnO single crystals were synthesized by high-pressure synthesis using a multi-anvil apparatus, as described elsewhere \cite{gil05}. Prior to the DS and IXS experiments, two untwinned single crystals selected from the same batch were mechanically made into the shape of a needle and etched down to 50 $\mu$m diameter with HCl in order to remove the damaged surface layer. The crystals were subsequently glued on a capillary and oriented with the $[001]$ direction of the cubic lattice perpendicular to the scattering plane. For the DS measurements, the samples were submitted to shutterless exposure with wavelength $\mathit{\lambda}$=0.715 \AA\ at the ID28 beamline of the European Synchrotron Radiation Facility (ESRF). Scattered intensities were measured using a PILATUS 2M detector, as described elsewhere \cite{dya16}. The rotation axis was set perpendicular to the beam and parallel to its polarization. The resolution was 0.1$^{\circ}$ per image and the exposure time was 0.6 s for each image. The data were analyzed using the CrysAlis software package and reconstructed using ESRF home-made software. Laue symmetry was applied to the data. The IXS experiment was performed at the ID28 beamline using a (999) Si monochromator line of wavelength $\mathit{\lambda}$=0.6968 \AA, corresponding to an energy resolution of 3.0 meV. Constant-${\bf Q}$ energy scans were collected in transmission mode with ${\bf Q} \perp [001]$. For both experiments, the sample temperature was controlled using a liquid-nitrogen cryostream apparatus. 


Lattice dynamical calculations for the cubic $Im\bar{3}$ phase of NaMnO were carried out using a shell model with pair-wise interionic interaction potential that includes short-range and long-range Coulomb terms. The parameters of the potential satisfy the conditions of static and dynamic equilibrium. The calculations, carried out using the current version of the DISPR software \cite{dispr}, include total energy calculations of the crystal structure, the phonon dispersion relation and the IXS cross section for each phonon mode. We computed the one-phonon IXS cross section for a given momentum ${\bf Q}$ and energy $\omega$ transfer using the expression:

\begin{equation}
\begin{split}
S({\bf Q},\omega)=A\sum_{{\bf q},j}\frac{\hbar}{2\omega({\bf q},j)}\left \{ n\left[\omega({\bf q},j)\right] +\frac{1}{2}\pm\frac{1}{2}\right\} \cdot  \\ \cdot \left|F_{j}(\textbf{Q})\right|^{2}\delta\left({\bf Q}-{\bf G}\mp\textbf{q}\right)\delta\left[\omega \pm \omega(\textbf{q},j)\right]
\end{split}
\end{equation}

where $A$ is a proportionality constant and the sum is extended over all phonon modes of wave vector ${\bf q}$, energy $\omega$ and branch $j$, $n \left[\omega({\bf q},j) \right] = \left\{ \exp\left[ \frac{\hbar\omega({\bf q},j)}{k_B T} \right] - 1 \right\}^{-1}$ is the Bose-Einstein distribution function for each mode, the sign $\pm$ indicates anti-Stokes and Stokes processes, respectively, ${\bf G} = {\bf Q} \mp {\bf q}$ is a reciprocal lattice vector and $F_{j}({\bf Q})$ is the dynamical structure factor:

\begin{equation}
\begin{split}
F_{j}(\textbf{Q})=\sum_{k=1}^N{f_{k}(\textbf{Q})\frac{\textbf{Q}\cdot{\bf e}({\bf q},j,k)}{\sqrt{m_k}}{\rm e}^{i\textbf{G}\cdot \textbf{R}_k}{\rm e}^{-W_k({\bf Q})}}
\end{split}
\end{equation}

where the sum is extended over all $N$ atoms in the unit cell, $f_k$, $m_k$, ${\bf e}({\bf q},j,k)$, ${\bf R}_k$ and $W_k$ are the form factor, mass, normalized eigenvector, vector position and Debye-Waller factor of $k$-th atom and ${\bf q}$ and $j$ are the phonon wave vector and branch, as above.

\section{Diffuse x-ray scattering results}

\begin{figure}[htpb]
\centering
\includegraphics[width=0.9\columnwidth]{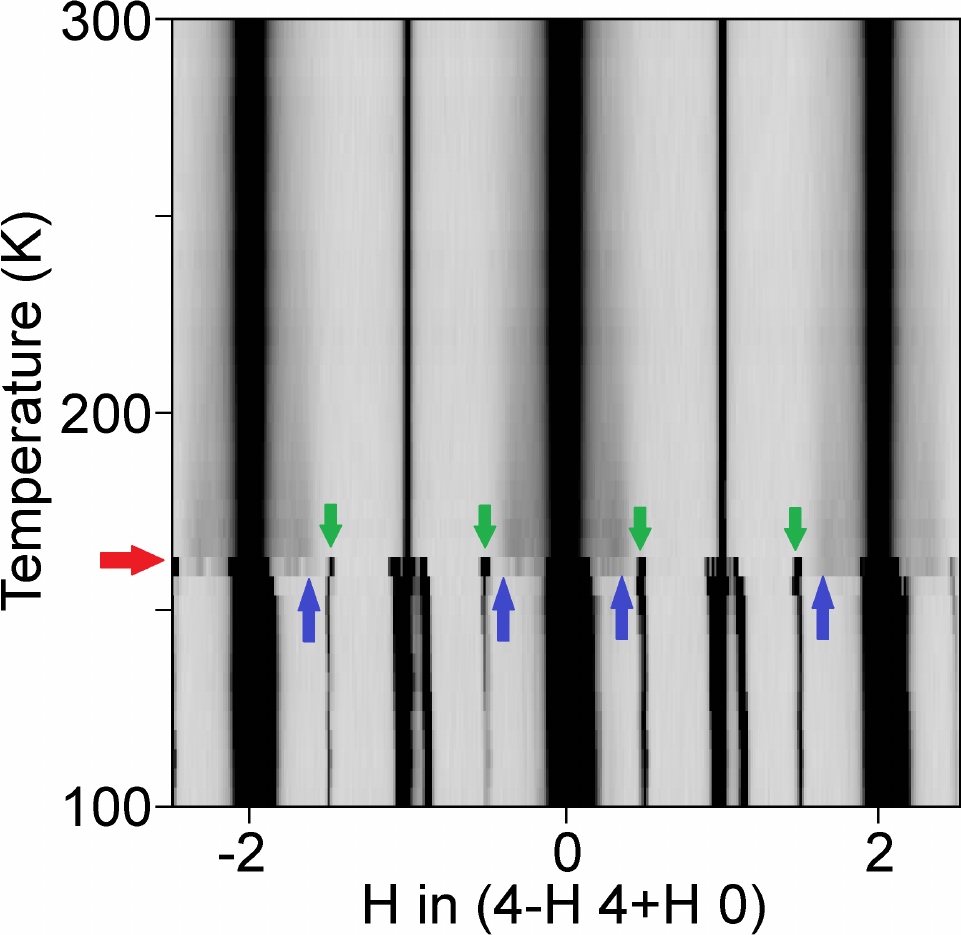}\caption{Temperature evolution of the DS intensity profile along the $\Sigma$ direction crossing the (440) peak. Blue arrows indicate the position of the features at ${\bf q}_{\rm DS} = (\frac{1}{3}\, \frac{1}{3}\, 0)$ that appear above 230 K and disappear upon cooling at the charge-ordering (CO) transition at $T_{\rm CO}$=175 K, indicated by the red arrow. At $T_{\rm CO}$, green vertical arrows mark the appearance of the superstructure reflections at ${\bf q}_{\rm CO} = (\frac{1}{2}\; \frac{1}{2}\; 0)$. The noise in the data below $T_{\rm CO}$ is attributed to the formation of twin domains in the CO monoclinic structure.}
\label{DS_cuts2}
\end{figure}

\begin{figure}[hb]
\centering
\includegraphics[width=\columnwidth]{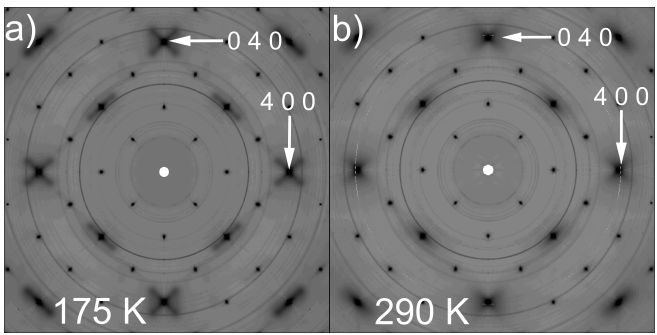}\caption{Detailed view of the reciprocal space maps of DS intensities near the (400) and (040) peaks at the CO transition temperature, $T_{\rm CO}$=175 K (a), and at room temperature (b).}
\label{DS_T}
\end{figure}

\begin{figure}[htpb]
\includegraphics[width=0.8\columnwidth]{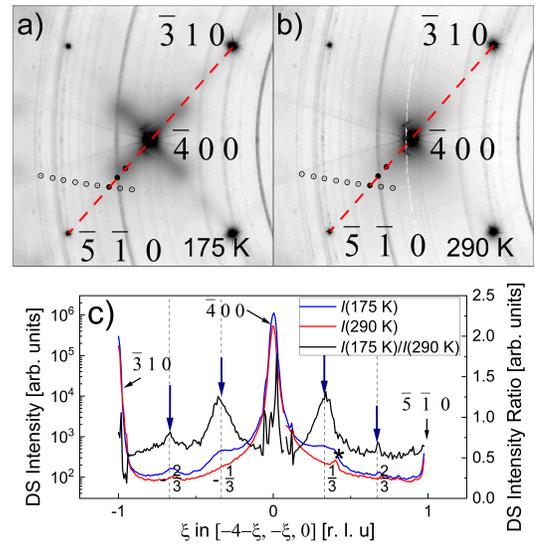} \caption{Color online. (a,b): reciprocal space DS intensity maps near the $\left(\bar{4}00\right)$ peak at $T_{\rm CO}$=175 K and at room temperature (290 K). At $T_{\rm CO}$, note the anomalous X-shaped features at ${\bf Q}={\bf G}_{hk0} \pm {\bf q}_{\rm DS}=(h\pm \frac{1}{3},k \pm\frac{1}{3},0)$. \textcolor{black}{Circles indicate the ${\bf Q}$ points measured in the IXS experiment by using the main analyzer (full circles) and the extra analyzers (open circles) (see also text and Fig. \ref{soft_phonon2})}. The broken red line indicates the direction of the intensity profiles shown in panel (c). (c): \textcolor{black}{DS intensity profiles at $T_{\rm CO}$=175 K and 290 K (red and blue lines) and ratio of these two profiles (black line)} along the $\Sigma$ direction crossing the $\left(\bar{4}00\right)$ peak, where the DS anomalous features are most visible. Blue arrows mark the wave vectors $\left(h\pm\frac{1}{3},k\pm\frac{1}{3},0\right)$ and $\left(h\pm\frac{2}{3},k\pm\frac{2}{3},0\right)$ where the anomalous DS intensities are observed. As in the DS data, quasi-elastic scattering is absent at room temperature. The asterisk indicates the polycrystal contribution to the diffraction pattern, visible as rings in panels (a,b).}
\label{CO} 
\end{figure}

By approaching gradually $T_{\rm CO}$ from room temperature, we expected to observe anomalous features in the DS data at the propagation vector ${\bf q}_{\rm CO}$ of the CO phase, which would reflect the incipient CO transition. \textcolor{black}{Surprisingly, we instead found anomalous features at completely different wave vectors ${\bf Q}=(h\pm\frac{1}{3},k\pm\frac{1}{3},0)$. As seen in Figs. \ref{DS_cuts2}, \ref{DS_T}a-b and \ref{CO}a-b, these features appear well above $T_{\rm CO}$ and become clearly visible already at $\approx$ 230 K; they} consist of X- or bar-shaped clouds visible especially around the most intense ($4m,0,0$) and ($2m+1,2n+1,0$) peaks, respectively, and equivalent peaks of the cubic $Im\bar{3}$ phase ($m$, $n$ integers). This observation is accompanied by a pronounced peak broadening along the $[110]$ and $[1\bar{1}0]$ directions. \textcolor{black}{The DS intensity profiles along these directions evolve continuously upon cooling down from room temperature. As seen in Fig. \ref{CO}c, the tail of the Bragg peak is progressively enhanced until well-defined satellite peaks appear at ${\bf Q}=(h\pm\frac{1}{3},k\pm\frac{1}{3},0)$ in the vicinity of $T_{\rm CO}$.} The second-order satellites at ${\bf Q}=(h\pm\frac{2}{3},k\pm\frac{2}{3},0)$ are also visible. This result indicates an incipient structural modulation with propagation vector ${\bf q}_{\rm DS}=(\frac{1}{3},\frac{1}{3},0)$. Upon further cooling, the intensity of the DS anomalies progressively increases until they abruptly disappear at $T_{\rm CO}$ (see Fig. \ref{DS_cuts2}), concomitant to the appearance of the satellite peaks of the CO structural modulation at ${\bf q}_{\rm CO}$. This indicates that the instability at ${\bf q}_{\rm DS}$ is suppressed by the CO instability at ${\bf q}_{\rm CO}$.

\section{Inelastic x-ray scattering results}

\begin{figure}[htpb]
\includegraphics[width=\columnwidth]{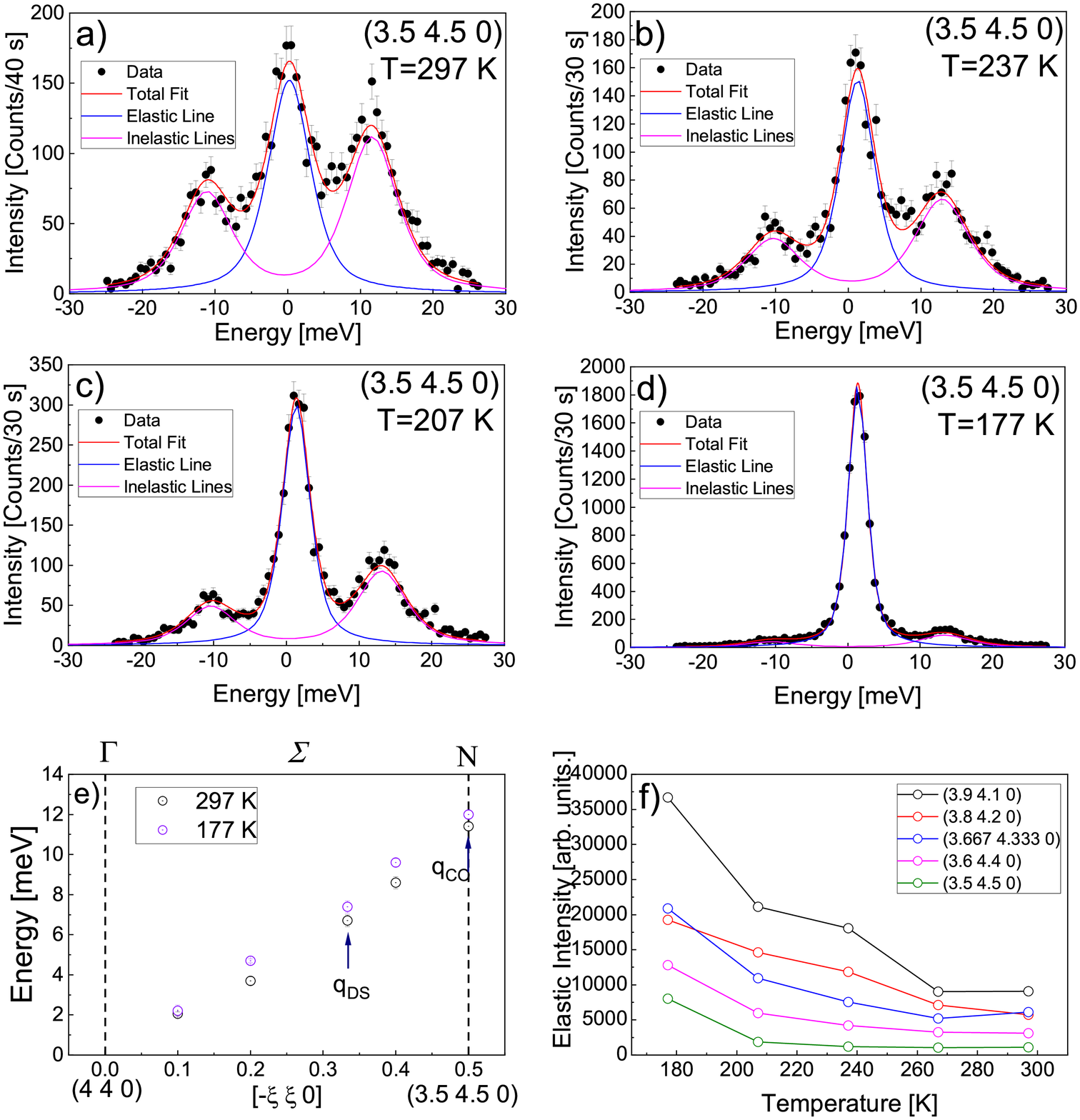} \caption{Color online. (a-d): Temperature evolution of the low-energy IXS spectra at ${\bf G}_{440} + {\bf q}_{\rm CO}=(3.5,4.5,0)$ down to 177 K, i.e. just above $T_{\rm CO}$. Note the increasing intensity of the elastic peak with decreasing $T$ and the Stokes and anti-Stokes acoustic modes. Solid lines are a fit of these 3 peaks, as explained in text. (e): Dispersion of the acoustic mode along $\Sigma$ at 297 and 177 K. Note a conventional mode hardening upon cooling, with no anomalies either at ${\bf q}_{\rm DS}$ or at ${\bf q}_{\rm CO}$. (f): temperature dependence of the elastic intensity at various $Q$-points along $\Sigma$. The intensity increases upon cooling and decreases by moving away from ${\bf G}_{440}$, except at ${\bf {q}}_{\rm DS}$, consistent with the X-shaped feature of the DS intensity in Fig. \ref{CO}a.}
\label{IXS_QE} 
\end{figure}

In order to unveil the dynamics of this latent instability, we measured IXS spectra at ${\bf Q}={\bf G}_{400}+{\bf q}=(4\pm \xi,0,0)$ and ${\bf Q}={\bf G}_{440}+{\bf q}=(4\pm\xi,4\pm\xi,0)$, \textit{i.e.} along the $[010]$ ($\Delta$) and $[110]$ ($\Sigma$) directions crossing the X- and bar-shaped DS anomalous features of Fig. \ref{CO}a. We focused on the low-energy $\pm 30$ meV range, which enabled us to investigate the quasi-elastic (QE) peak and the longitudinal and mixed longitudinal-transverse acoustic phonons. Data were taken at 177 K, i.e. just above $T_{\rm CO}$, and at 207, 237 and 297 K. $T_{\rm CO}$ was precisely determined to be 175 K by cooling the sample until the satellite peaks of the CO phase appeared and by reducing the cooling rate until the thermal hysteresis disappeared. Representative spectra are shown in Fig. \ref{IXS_QE}. The mode energies were determined by fitting the experimental phonon peaks using pseudo-Voigt functions. The Stokes and anti-Stokes lines of the acoustic phonons were fitted simultaneously by imposing the constraint set by the Bose factor for the intensities. Elastic lines were analyzed using a sum of Gaussian and Lorentzian line shapes to account for the contributions of the experimental resolution and of quasi-elastic scattering, respectively.

The main result is an enhanced intensity of the QE peak upon cooling down to $T_{\rm CO}$, similar to the case of incommensurate orbital ordering transition in the related compound \CaMnO \cite{sou16}. Interestingly, in the present case, the QE intensity progressively decreases by moving away from the above $(hk0)$ peaks, except for a sizable enhancement at ${\bf Q}={\bf G}_{hk0}+{\bf q}_{\rm DS}$ (see Fig. \ref{IXS_QE}f), consistent with the behavior of the DS intensity profile of Fig. \ref{CO}c. Upon cooling, no anomaly is seen in the mode dispersion at ${\bf q}_{\rm DS}$ (see Fig. \ref{IXS_QE}e), which is again consistent with the DS result that the instability at ${\bf q}_{\rm DS}$ is suppressed at $T_{\rm CO}$.

\begin{figure}[htpb]
\centering \includegraphics[width=0.9\columnwidth]{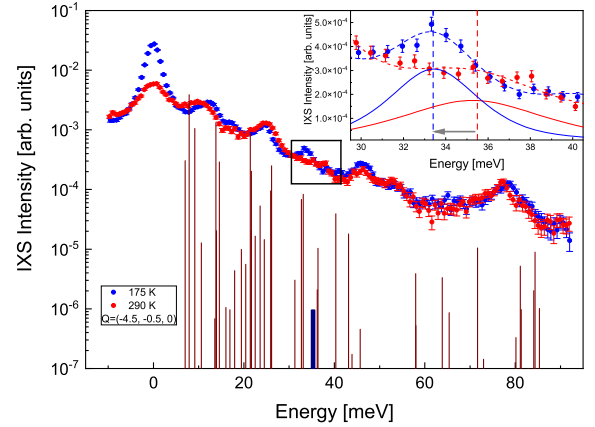}

\caption{Color online. IXS spectra taken at 295 K (red points) and $T_{\rm CO}$=175 K (blue points) at the CO wave vector ${\bf Q}^{*}=(-4.5, -0.5, 0)$. Vertical bars indicate the calculated energy and intensity of the A$_g$ and B$_u$ modes (see Table \ref{E_I}). The A$_g$ mode calculated at 35.3 meV, indicated by a thick blue bar, is attributed to the soft mode highlighted in the black frame and in the inset. The polarization of this A$_g$ mode is given in Table \ref{Ag}. Inset: detail of the softening region. Solid red and blue lines are a fit of the mode at 295 and 175 K, respectively. The vertical broken lines and the arrow indicate the $\approx$2 meV energy shift of the mode. Broken lines are a fit of the two spectra including the contribution of all modes.}
\label{soft_phonon} 
\end{figure}

In the search for anomalous modes at the verge of the CO transition, in a second IXS experiment we measured the spectra along the same $\Delta$ and $\Sigma$ directions as before in an extended energy range up to 90 meV, which enabled us to probe all modes. To look for anomalies, ideally one should follow the changes in the modes across $T_{\rm CO}$. In practice, this is hindered by the formation of twinned domains in the monoclinic CO phase below $T_{\rm CO}$. Still, the thermodynamic fluctuations of the CO order parameter near the transition can be sufficiently strong to produce measurable structural anomalies. It is expected that, slightly above $T_{\rm CO}$, these fluctuations produce a measurable rounding in the temperature-dependence of the critical softening of the ${\bf q}_{\rm CO}$ phonon in the cubic phase.  

\begin{figure}[htpb]
\centering
\includegraphics[width=\columnwidth]{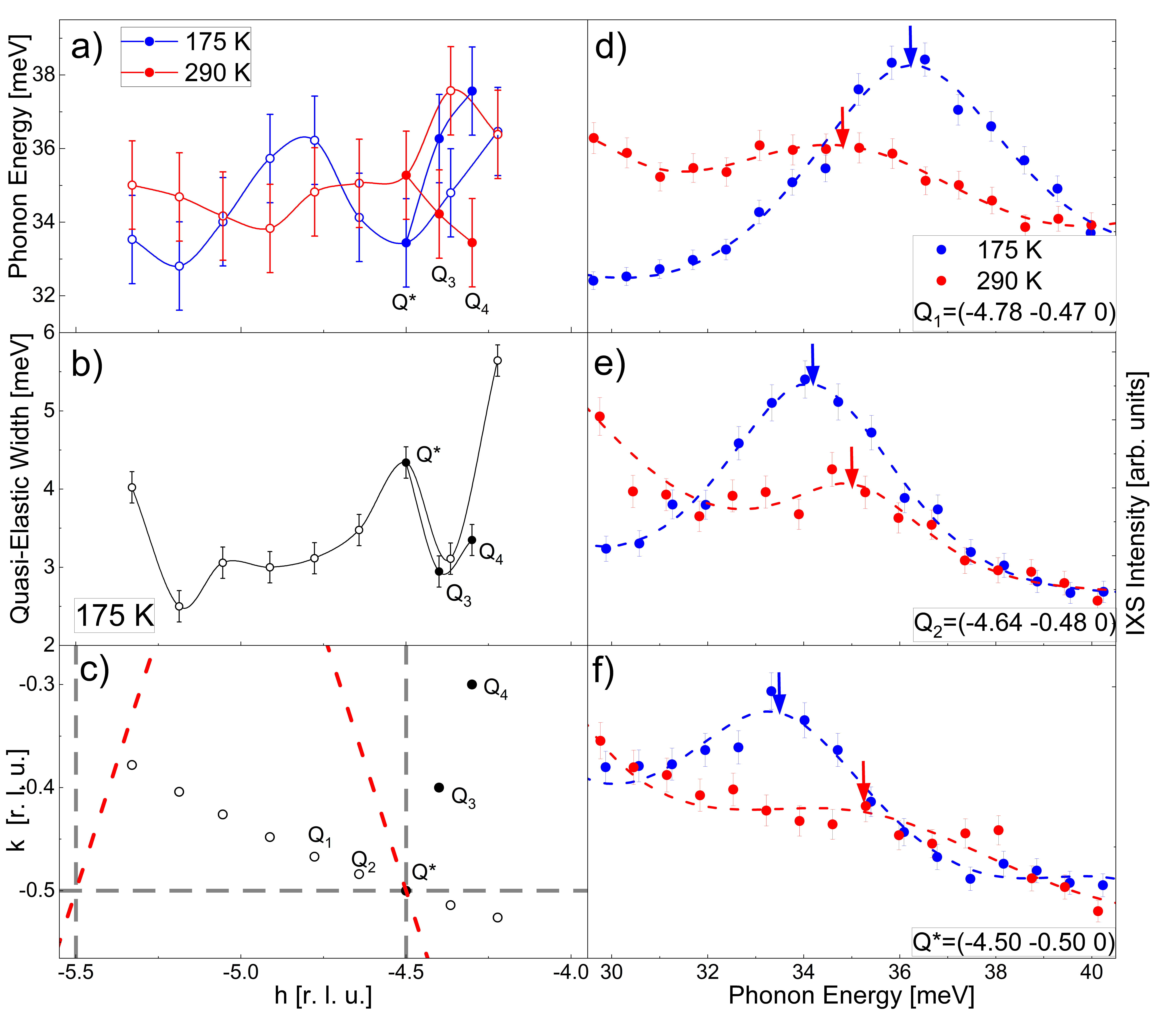}
\caption{Color online. (a): IXS dispersion of the soft mode of Fig. \ref{soft_phonon} taken at 290 K and $T_{\rm CO}$=175 K along the $[\bar 1 1 0]$ ($\Sigma$) direction and along the arc indicated in panel (c). Full and open symbols are points measured by the main and extra analyzers, respectively (see text). Lines are a guide to the eye. Note a pronounced softening at $T_{\rm CO}$ by approaching the wave vector ${\bf Q}^{*}={\bf G}_{400}-{\bf q}_{\rm CO}$. (b): ${\bf Q}$-dependence of the width of the quasi-elastic peak at $T_{\rm CO}$. Note the peak broadening at ${\bf Q}^{*}$. Full and open symbols are as in (a) and (c). (c): Location in the ($hk0$) plane of all measured ${\bf Q}$-points in reciprocal lattice units [r.l.u.]. Broken red and gray lines indicate the Brillouin zone boundary of the cubic unit cell and of the monoclinic supercell, respectively. (d-f): detail of the spectra at ${\bf Q}^{*}=(-4.5, -0.5, 0)$, ${\bf Q}_2=(-4.64, -0.48, 0)$ and ${\bf Q}_1=(-4.78, -0.47, 0)$. Broken lines are a fit of the full spectrum to which all modes contribute. Arrows indicate peak position.}
\label{soft_phonon2}
\end{figure}

This expectation is confirmed by the spectra taken along $[\bar 1 1 0]$ ($\Sigma$) (see Fig. \ref{soft_phonon}). Upon cooling, note a pronounced 6\% softening of an optical mode around 35 meV at the critical wave vector ${\bf Q}^{*}={\bf G}_{\bar{4}00}-{\bf q}_{\rm CO}=(-4.5,-0.5,0)$. The critical behavior of the softening of this ${\bf q}_{\rm CO}$ phonon is evident by analyzing its dispersion along $\Sigma$. Additional ${\bf Q}$ points collected by 8 extra analyzers located in the scattering plane or slightly above give further evidence of this softening. These additional points enable us to reconstruct the dispersion curve along an arc in the $(hk0)$ plane crossing ${\bf G}_{\bar{4}00}$, as indicated in Figs. \ref{CO}a,b and \ref{soft_phonon2}a-c. In conclusion, Fig. \ref{soft_phonon2} clearly shows a progressive mode softening as the critical wave vector ${\bf Q}^{*}$ is approached while, away from ${\bf Q}^{*}$, the mode displays a conventional hardening upon lowering temperature. The magnitude of the softening is noteworthy, considering that it occurs in the fluctuation region of the cubic phase and that, in the absence of anomalies, a mode hardening should instead occur.

The ${\bf Q}$-dependence of the QE peak in Fig. \ref{soft_phonon2}b further indicates that the softening is correlated with incipient short-range structural correlations at $T_{\rm CO}$. Knowing that this peak is well described by a Lorentzian function \cite{mas16}, we focus on the changes in the peak intensity and width observed at $T_{\rm CO}$. The energy dependence of the peak carries significant information on the lattice dynamics because the peak is broader than the instrumental resolution. We exclude that the broadening is due to the contribution of low-energy acoustic phonons, for these phonons are located at higher energies. Our analysis of the peak data as a function of ${\bf Q}$ shows a significant peak broadening at ${\bf Q}^{*}$ (see Fig. \ref{soft_phonon2}b), which corroborates the picture of critical behavior of the lattice dynamics at ${\bf q}_{\rm CO}$.

\section{Phonon calculations results}
We should now try to single out the driving force of the lattice softening by identifying the symmetry of the soft phonon. In the high-temperature cubic $Im\bar{3}$ phase, the small point group is cubic $T_h$ at $\Gamma$ and monoclinic $C_{2h}$ at ${\bf q}_{\rm CO}={\rm N}$, where the softening is observed, so the symmetry of the measured phonons at ${\bf q}_{\rm CO}$ is either A$_{g}$ or B$_{u}$. In NaMnO, there are 8 modes of the former symmetry and 26 of the latter. We then carried out lattice dynamical calculations on the high-temperature cubic $Im\bar{3}$ phase. The room temperature structural data used as input for the calculations are taken from \cite{mar73}. In Table \ref{E_I} below we report the calculated energies and intensities of the 34 A$_g$ or B$_u$ modes, symmetric with respect to $\sigma_z$, at the wave vector ${\bf Q}^{*}$ measured in the IXS experiment. We do not consider the remaining 26 B$_g$ or A$_u$ modes, for they are antisymmetric with respect to $\sigma_z$ and therefore not measured in the present IXS geometry. In Table \ref{Ag}, we give the eigenvector of the relevant A$_g$ mode of calculated energy 35.3 meV, attributed to the soft mode observed by IXS.

\begin{figure}[htpb]
\centering \includegraphics[width=\columnwidth]{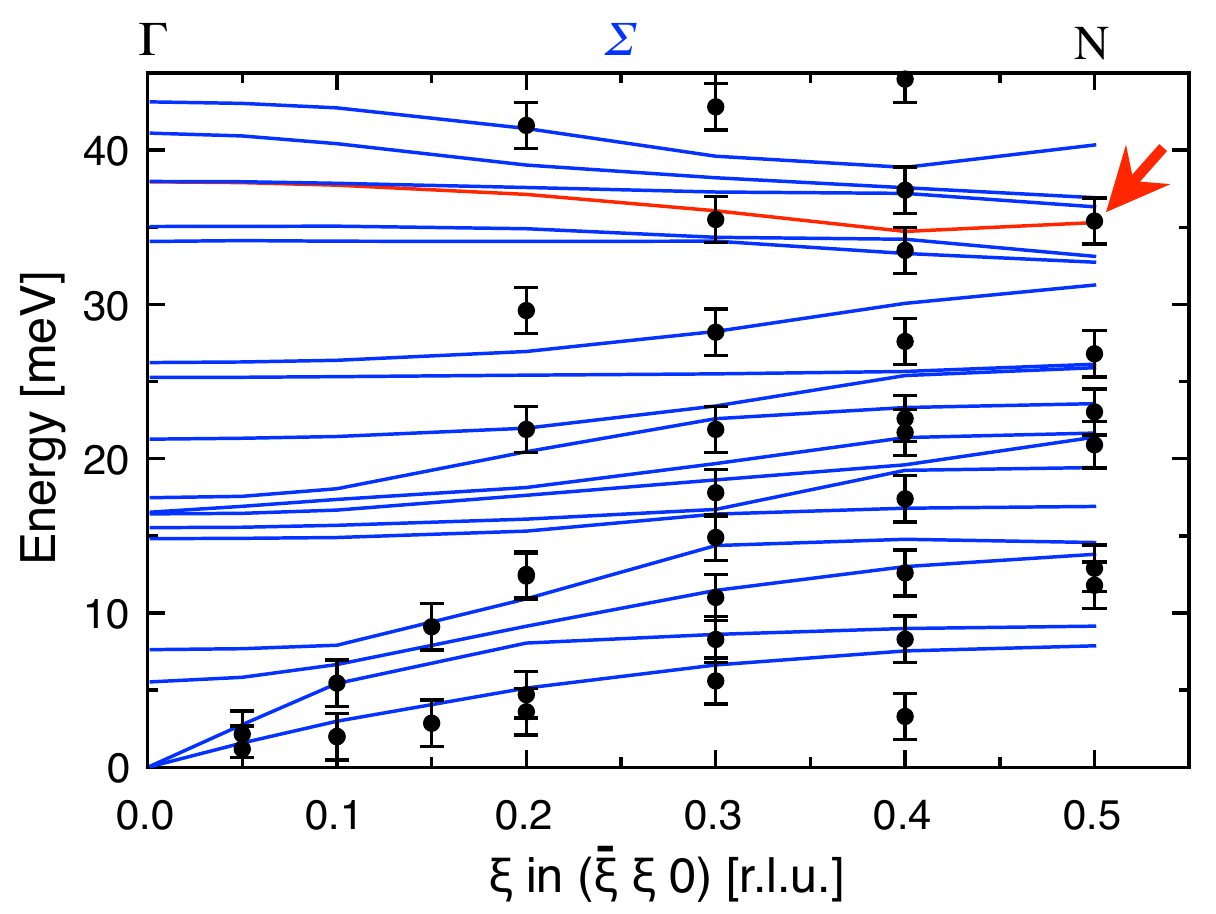}
\caption{Experimental (points) and calculated (lines) phonon dispersion along the $[\bar{1} 1 0]$ ($\Sigma$) direction in reciprocal lattice units [r.l.u.]. The experimental points have been measured by IXS at 290 K (see main text). The red line indicates the mode attributed to the soft mode observed experimentally at the zone boundary ${\rm N}=(-\frac{1}{2},\frac{1}{2},0) = {\bf q}_{\rm CO}$ marked by a red arrow. At this point the mode symmetry is A$_{g}$. See in Table \ref{E_I} the list of calculated mode energies and intensities at N and in Table \ref{Ag} the calculated polarization of the soft A$_{g}$ mode.}
\label{dispersion} 
\end{figure}

\begin{figure}[htpb]
\includegraphics[width=\columnwidth]{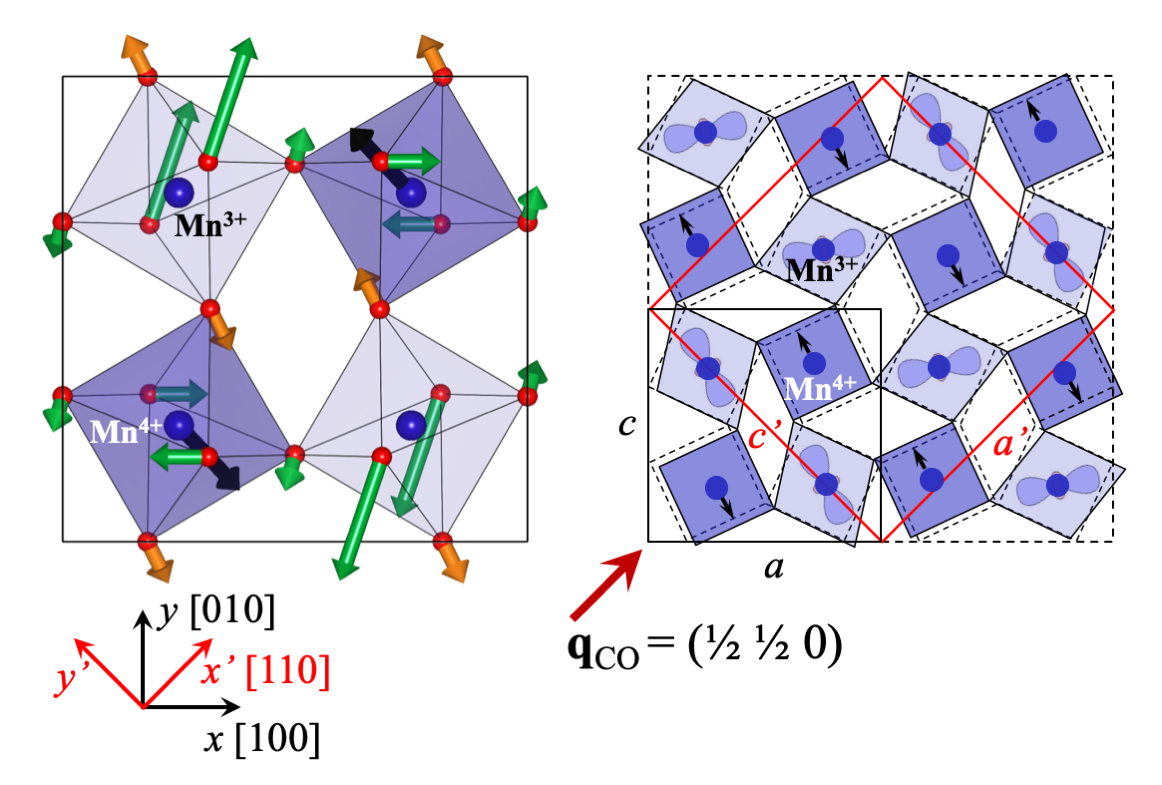}
\caption{Color online. Left: calculated polarization of the A$_{g}$ 35.3 meV mode at ${\bf q}_{\rm CO}=(\frac{1}{2} \frac{1}{2} 0)$ in the parent cubic $Im\bar 3$ symmetry, attributed to the soft phonon observed by IXS. Arrows indicate the displacements (in scale) of the octahedral $B$-site Mn and O atoms. The values of the displacements are reported in Table \ref{Ag} for all atoms. Dark and light blue colors denote the two distinct $4e$ and $4f$ B-sites in the A$_{g}$ symmetry (see Table \ref{Ag}). Right: Schematic structural modulation described by the propagation vector ${\bf q}_{\rm CO}$ in the monoclinic $C2/m$ symmetry, concomitant to the Mn$^{3+}$/Mn$^{4+}$ charge ordering at $T_{\rm CO}$ (adapted from \cite{pro14}). The unit cell of the parent $Im\bar 3$ structure and the supercell are indicated by thick solid black and red lines, respectively. Thin broken and solid black lines represent the $B$-site MnO$_6$ octahedra in the parent and distorted structures, respectively. Note the match between the pattern of the atomic displacements of the A$_{g}$ mode and the structural modulation, in particular the zig-zag displacement of the Mn$^{4+}$ ions (black arrows) and the Jahn-Teller elongation of the octahedra of the Mn$^{3+}$ ions (orange arrows) leading to a zig-zag orientation of the 3$d_{z^2}$ orbitals of these ions.}
\label{JT_phonon}
\end{figure}

The reliability of the calculations is seen in Fig. \ref{dispersion}, where one notes the agreement between experimental and calculated dispersions along the representative $[\bar110]$ direction within the experimental resolution. Hereafter, we focus on the above 34 A$_{g}$ and B$_{u}$ modes at the ${\bf Q}^{*}$ wave vector of interest. In Fig. \ref{soft_phonon}, we note the agreement between experimental and calculated positions and intensities of these modes at ${\bf Q}^{*}$ (see Table \ref{E_I} for a full list of the calculated values). Thus, we are confident that the shell model employed in the calculations is reliable to identify the polarization of the soft mode. 

Within the energy resolution of the IXS experiment, $\Delta E$=3.0 meV, three of the calculated modes, two with B$_{u}$ symmetry and one with A$_g$ symmetry, have energies of $35 \pm1.5$ meV at ${\bf Q}^{*}$ (see Table \ref{E_I}). However, only the polarization of the $A_g$ mode, of calculated energy 35.3 meV, matches the displacement of the Mn and O atoms governing the structural modulation in the CO phase previously reported \cite{pro14}. This is understood using the straightforward symmetry argument that A$_g$ is the totally symmetric representation of the $C_{2h}$ small group of the modulated structure. The match between distortion pattern of the phonon and structural modulation is evident in Fig. \ref{JT_phonon}. Note, in particular, the JT-like elongation of half of the MnO$_{6}$ octahedra in the pristine cubic cell along the apical $[100]$ or $[010]$ direction, leading to a zig-zag pattern along the $[110]$ direction of the ${\bf q}_{\rm CO}$ propagation vector. This leads to the formation of the $\sqrt{2}\times\sqrt{2}$ supercell previously reported in the monoclinic CO phase \cite{pro14}. The observed JT distortion of these octahedra is as expected for the JT-active Mn$^{3+}$ ions, consistent with the scenario of Mn$^{3+}$/Mn$^{4+}$ order previously reported. The remaining octahedra, occupied by the Mn$^{4+}$ ions in the CO phase, exhibit a modest distortion with negligible volume change, for these ions are not JT active. They instead exhibit an alternated displacement along the $[110]$ direction to accommodate the elongation of the Mn$^{3+}$ octahedra. In summary, the CO modulated structure arises from a critical slowing down of the Jahn-Teller 35.3 meV A$_g$ phonon.

\section{Conclusions}
In conclusion, we directly observed the critical lattice dynamics at the CO Verwey transition in the mixed-valence compound NaMnO by means of DS and IXS. This critical dynamics consists of a phonon softening at the propagation vector ${\bf q}_{\rm CO}$ of the CO modulated structure. Remarkably, the softening is sizable already in the thermodynamic fluctuation region of the cubic phase slightly above $T_{\rm CO}$. Lattice dynamical calculations enabled us to attribute the softening to a A$_{g}$ optical mode coupled to the JT dynamics of the MnO$_{6}$ octahedra in presence of an incipient Mn$^{3+}$/Mn$^{4+}$ charge disproportionation. The match between calculated polarization of the soft phonon and experimental distortion pattern of the CO phase shows that the JT instability is the microscopic mechanism of the transition. The data also unveil a competing structural instability with a distinct commensurate propagation vector appearing at temperatures much higher than $T_{\rm CO}$ and suppressed by the CO instability. Further studies using experimental probes of the short-range crystal structure, such as extended X-ray absorption fine-structure (EXAFS), may further elucidate the long-standing controversy on the scenarios of static charge ordering vs. dynamical charge fluctuations governing the physics of mixed-valence transition-metal oxides including magnetite, manganites and superconducting cuprates.

\begin{acknowledgments}
The authors gratefully acknowledge the ESRF for the beamtime allocated to experiments HC1405 and HC2499 and financial support provided by the FAPESP (projects 2015/21206-1, 2013/27097-4 and 2013/07296-2), CAPES (Finance Code 001, project CAPES-COFECUB 88887.130195/2017-01), CNPq (projects 426965/2018-3 and 311462/2017-0) and the Indian National Science Academy for financial support of an INSA Senior Scientist position to S.L.C.
\end{acknowledgments}

\section{Appendix}

\begin{table}[ht]
\centering
\caption{Calculated energy and intensity of the 34 A$_g$ or B$_u$ phonon modes measured in the IXS experiment at the critical wave vector ${\bf q}_{\rm CO} = (\frac{1}{2}\; \frac{1}{2}\; 0)$ of the charge-order Verwey transition. The calculated energy and intensity values are plotted graphically as vertical bars in Figure \ref{soft_phonon}. In bold the A$_g$ mode attributed to the soft mode observed in the IXS experiment (see Figures \ref{soft_phonon} and \ref{soft_phonon2}).}
\renewcommand{\arraystretch}{1.2}
\begin{tabular}{c@{\hskip 0.5cm}c@{\hskip 0.5cm}c}
\hline
\hline
{\bf Energy} [meV] & {\bf Calculated intensity} [arb. units] & {\bf Symmetry} \\ 
\hline
\hline
7.9  & 3.89 $\times 10^{-3}$ & A$_g$ \\ \hline
9.2  & 1.05 $\times 10^{-3}$ & B$_u$ \\ \hline
13.8 & 1.32 $\times 10^{-3}$ & B$_u$ \\ \hline
14.6 & 2.90 $\times 10^{-4}$ & A$_g$ \\ \hline
16.9 & 9.67 $\times 10^{-7}$ & A$_g$ \\ \hline
19.4 & 9.90 $\times 10^{-6}$ & B$_u$ \\ \hline
21.4 & 8.75 $\times 10^{-4}$ & A$_g$ \\ \hline
21.7 & 2.01 $\times 10^{-4}$ & B$_u$ \\ \hline
23.6 & 5.30 $\times 10^{-5}$ & A$_g$ \\ \hline
25.9 & 9.47 $\times 10^{-5}$ & B$_u$ \\ \hline
26.1 & 2.51 $\times 10^{-4}$ & A$_g$ \\ \hline
31.3 & 3.02 $\times 10^{-6}$ & B$_u$ \\ \hline
31.6 & 0.00                  & A$_g$ \\ \hline
32.7 & 6.80 $\times 10^{-5}$ & B$_u$ \\ \hline
33.1 & 8.30 $\times 10^{-5}$ & B$_u$ \\ \hline
\textbf{35.3} & \textbf{9.59} $\times {\bf 10^{-7}}$ & \textbf{A}$_{\bf g}$ \\ \hline
36.3 & 1.03 $\times 10^{-5}$ & B$_u$ \\ \hline
36.9 & 0.00                  & A$_g$ \\ \hline
40.4 & 3.93 $\times 10^{-5}$ & B$_u$ \\ \hline
42.2 & 0.00                  & A$_g$ \\ \hline
43.1 & 1.80 $\times 10^{-5}$ & A$_g$ \\ \hline
45.7 & 4.56 $\times 10^{-7}$ & B$_u$ \\ \hline
57.9 & 3.89 $\times 10^{-6}$ & B$_u$ \\ \hline
58.0 & 5.22 $\times 10^{-7}$ & A$_g$ \\ \hline
63.9 & 3.33 $\times 10^{-6}$ & B$_u$ \\ \hline
65.4 & 8.63 $\times 10^{-7}$ & A$_g$ \\ \hline
71.7 & 1.05 $\times 10^{-5}$ & B$_u$ \\ \hline
72.9 & 1.43 $\times 10^{-7}$ & A$_g$ \\ \hline
80.2 & 3.31 $\times 10^{-7}$ & B$_u$ \\ \hline
81.1 & 5.19 $\times 10^{-6}$ & A$_g$ \\ \hline
82.9 & 8.25 $\times 10^{-9}$ & A$_g$ \\ \hline
84.1 & 2.02 $\times 10^{-6}$ & B$_u$ \\ \hline
84.4 & 8.89 $\times 10^{-6}$ & A$_g$ \\ \hline
85.3 & 1.01 $\times 10^{-6}$ & B$_u$ \\ \hline
\end{tabular}
\label{E_I} 
\end{table}

\begin{table*}[ht]
\caption{Calculated atomic displacements, $u_x$, $u_y$ and $u_z$, along the $x$-, $y$-, and $z$-directions for the 35.3 meV A$_g$ phonon. Atomic site label, Wyckoff position and symmetry are given for both parent $Im\bar3$ and distorted $C/2m$ structures. Atomic coordinates are given in reduced lattice units. The structural data of the $Im\bar3$ structure are taken from \cite{mar73}, with unit cell parameter $a=7.3036$ \AA. The atomic coordinates of the $C/2m$ structure are given by choosing a unit cell with unique axis $z$ and origin at the centre, corresponding to the $I112/m$ space group \cite{ITC16}.}
\renewcommand{\arraystretch}{1.2}
\begin{tabular}{c c c c c c c c c c c c}
\hline
\hline
\multicolumn{6}{c}{{\bf Atomic site}} & \multicolumn{3}{c}{{\bf Atomic coordinates}} & \multicolumn{3}{c}{{\bf Atomic displacements} [arb. units]} \\
\hline
\multicolumn{3}{c}{\thead{Parent\\$Im\bar3$ structure}} & \multicolumn{3}{c}{\thead{Distorted\\$C/2m$ structure}} & $x$ & $y$ & $z$ & $u_x$ & $u_y$ & $u_z$ \\
\hline
\hline
Na & $2a$ & $m\bar 3$                                                  & Na & $2a$ & $2/m$ & 0   & 0   &  0  & 0 & 0 & 0 \\
\hline
\multirow{3}{*}{MnA'} & \multirow{3}{*}{$6b$} & \multirow{3}{*}{$mmm$} & \verb' 'MnA'1\verb' ' & $2b$ & \verb' '$2/m$\verb' ' & 0 & 0 & 1/2 & 0 & 0 & 0 \\
{}                   & {}                    & {}                      & MnA'2 & $2c$ & $2/m$ & 1/2 & 0 & 0 & $-0.53$ & $0.39$ & 0 \\
{}                   & {}                    & {}                      & MnA'3 & $2d$ & $2/m$ & 0 & 1/2 & 0 & $-0.03$ & $0.65$ & 0 \\
\hline
\multirow{2}{*}{MnB} & \multirow{2}{*}{$8c$} & \multirow{2}{*}{$\bar3$} & MnB1 & $4e$ & $\bar1$ & 1/4 & 1/4 & 1/4 & \verb' '$0.32$\verb' ' & \verb' '$-0.34$\verb' ' & \verb' '$-0.19$\verb' ' \\
{}                   & {}                    & {}                       & MnB2 & $4f$ & $\bar1$ & 1/4 & 1/4 & 3/4 & 0 & 0 & 0 \\
\hline  
\multirow{4}{*}{O} & \multirow{4}{*}{$24g$} & \multirow{4}{*}{$m$} & O1 & $4i$ & $m$ & \verb' '0.3132\verb' ' & \verb' '0.1828\verb' ' & 0 & $0.31$ & 0 & 0 \\
{}                   & {}                    & {}                  & O2 & $4i$ & $m$ & 0.6868 & 0.1828 & 0 & $0.22$ & $0.65$ & 0 \\
{}                   & {}                    & {}                  & O3 & $8j$ & $ 1 $ & 0.1828 & 0 & 0.3132 & $0.11$ & $-0.21$ & $0.06$ \\
{}                   & {}                    & {}                  & O4 & $8j$ & $ 1 $ & 0 & 0.3132 & \verb' '0.1828\verb' ' & $-0.05$ & $-0.19$ & $-0.09$ \\

\hline
\hline
\end{tabular}
\label{Ag} 
\end{table*}

\bibliography{namno_ixs}

\end{document}